\documentclass[
  aps,
  prb,
  twocolumn,
  superscriptaddress,
  amsmath,
  amssymb,
  floatfix,
  10pt
]{revtex4-2}

\usepackage{graphicx}
\usepackage{bm}
\usepackage[T1]{fontenc}
\usepackage[utf8]{inputenc}
\usepackage{adjustbox}
\usepackage[table]{xcolor}
\usepackage{xr-hyper}
\usepackage{hyperref}
\usepackage{tabularx}
\usepackage{makecell}
\usepackage{array}
\usepackage{xcolor}

\hypersetup{
  colorlinks=true,
  citecolor=blue,
  urlcolor=blue,
  linkcolor=blue
}

\begin{document}

\title{A General Algorithm for Minimal Uncoupled Stress--Strain Calculations of Elastic Constants}

\author{S.~I.~Simak}
\thanks{Corresponding author, sersi78@liu.se}
\affiliation{
Department of Physics, Chemistry, and Biology, Link\"oping University, 58183 Link\"oping, Sweden
}
\author{Utkarsh Singh}
\affiliation{
Department of Physics, Chemistry, and Biology, Link\"oping University, 58183 Link\"oping, Sweden
}

\date{\today}

\begin{abstract}
We have developed a general algorithm for constructing the minimum number of strain configurations required to extract second-order elastic constants from first-principles stress--strain calculations without algebraic coupling between the constants targeted for extraction. The method operates on the sparsity structure of the symmetry-reduced elastic stiffness matrix, represented by an auxiliary binary matrix, and generates admissible strain patterns algorithmically rather than prescribing separate strain sets for individual crystal symmetries. Each independent elastic constant is obtained from a single stress component, while additional stress relations generated by the same deformation provide internal consistency checks. Because the six-dimensional strain space contains only $2^6-1=63$ non-empty strain patterns, the minimality of the resulting sets can be verified exactly for every elastic-symmetry type considered. The construction reproduces established high-efficiency strain sets in cases where they are minimal and yields improved sets for symmetry classes in which previously tabulated schemes require additional configurations. Because the formulation is based on stress–strain derivatives, it can also be readily extended to finite temperatures using consistently sampled ensemble-averaged thermodynamic stresses.
Explicit strain sets are derived for all elastic-symmetry types spanning the 230 crystallographic space groups, and their accuracy and computational efficiency are benchmarked against established approaches.
\end{abstract}

\maketitle

\section{Introduction}

Elastic constants quantify the linear mechanical response of crystalline solids and provide fundamental input for assessing elastic stability, acoustic propagation, lattice dynamics, thermal stress, and mechanically driven phase behavior. First-principles elastic tensors are therefore important in both fundamental condensed-matter studies and materials modeling, with applications ranging from seismic-wave propagation in the Earth's crust and core \cite{Belonoshko2022,Mattesini2010} and acoustic transport in solids \cite{Belonoshko2019} to the design of piezoelectric devices \cite{Tasnadi2010}. Their accurate extraction becomes especially important when the response is required as a function of pressure, volume, temperature, or structural state.
For dynamic mechanical loading and damping problems, the elastic constants also enter the propagation and attenuation of mechanical waves, including the initial response of solids under high-strain-rate deformation and cyclic loading \cite{Steneteg2013,Damping}.

Among the available first-principles approaches, elastic constants can be obtained from free-energy--strain relations, stress--strain relations, or force-constant-based lattice-dynamical methods. The latter can be used at both zero and finite temperature, for example within the temperature-dependent effective potential (TDEP) framework \cite{Hellman2011,Hellman2013,Knoop2024}, but may require demanding convergence of the underlying calculations with respect to computational parameters \cite{Shulumba2015}. Free-energy and stress-based approaches therefore remain particularly useful for practical materials calculations. At $T=0$~K, the free-energy route reduces to a ground-state energy calculation and is conceptually straightforward. At finite temperature, however, the direct time average of the internal energy from molecular dynamics is not the Helmholtz free energy; entropy contributions must be accounted for explicitly. The free energy can nevertheless be reconstructed using appropriate methods \cite{Mosyagin2016}. A complementary route is to obtain the thermodynamic stress directly from first-principles calculations and molecular dynamics and determine the elastic response from its derivative with respect to strain.

The practical efficiency of the stress--strain approach depends critically on the choice of strain configurations. Because each stress component is generally a linear combination of several strain components, a single strain configuration can cause multiple elastic constants to enter the same stress equation.
Conventional schemes therefore either require several separately strained calculations or solve a coupled system of equations. For low-symmetry crystals, where many elastic constants are independent, this distinction becomes particularly important. It is desirable to construct strain configurations that exploit the sparsity imposed by crystal symmetry, isolate individual elastic constants during extraction, and use the smallest possible number of first-principles calculations.

The aim of the present work is to turn this construction into a general algorithm. We define an \emph{uncoupled} strain configuration as one for which every stress component used to extract an elastic constant contains only that target constant. We then represent the symmetry-allowed zero/nonzero structure of the stiffness tensor by an auxiliary binary matrix and use it to generate admissible strain patterns. The resulting procedure is deterministic, transparent, and independent of a manually tabulated strain set for any particular symmetry. Because the strain space has only six independent components, the minimality of the generated sets can be verified exactly by exhaustive enumeration of all non-empty strain patterns.

Before introducing the construction, we establish the strain and stress measures used throughout the paper and clarify the connection between the thermodynamic stress and the stresses obtained from first-principles calculations. We then derive the algorithm, illustrate it for a tetragonal symmetry, tabulate the resulting strain sets for all elastic-symmetry types, and benchmark the method against established stress--strain schemes at zero and finite temperature.

\begin{figure*}
    \centering
    \includegraphics[width=0.9\linewidth]{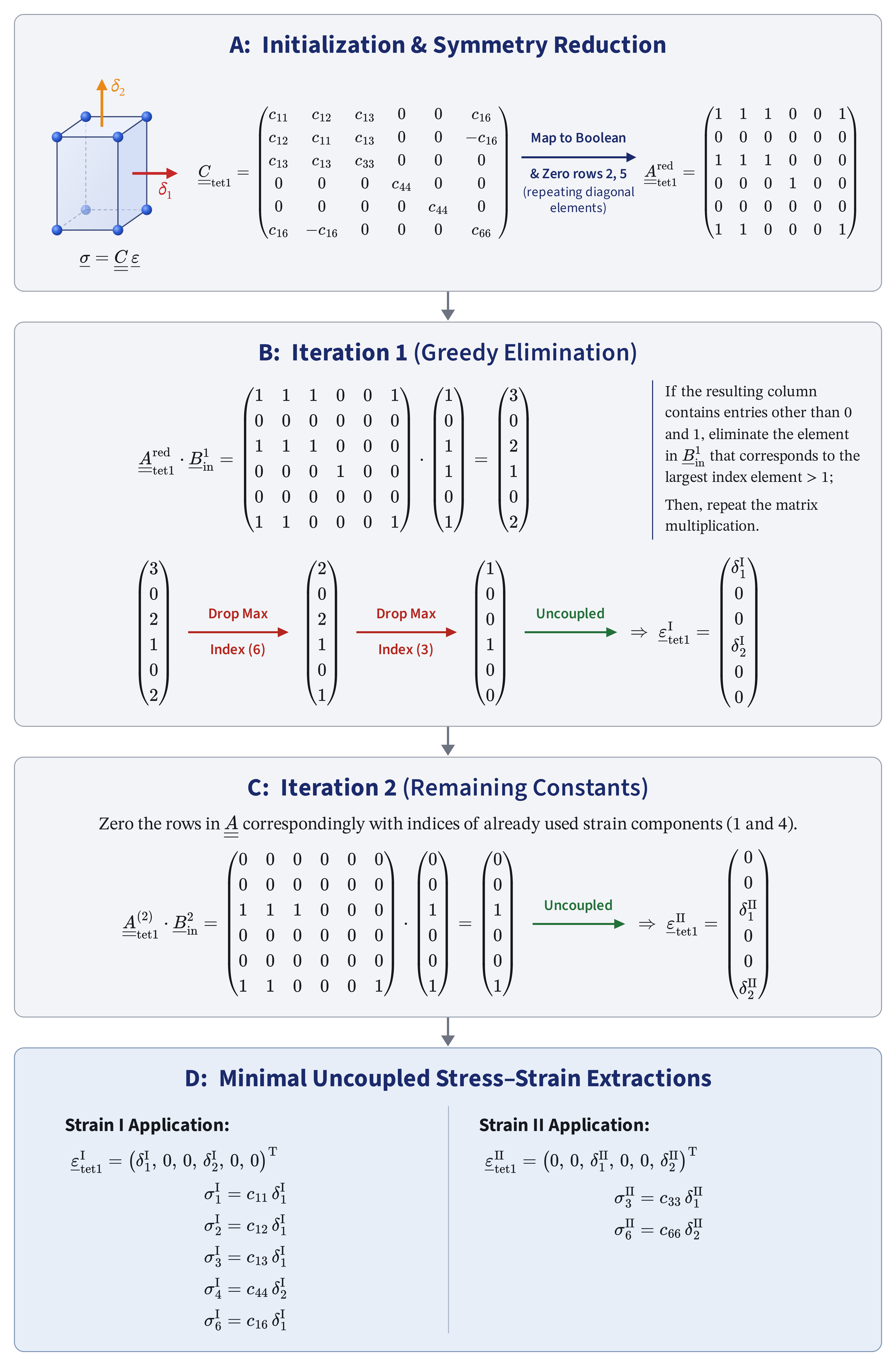}
    \caption{Conceptual Illustration of the minimal and uncoupled stress--strain scheme presented in this work for the example of Tetragonal-I crystal symmetry. 
    }
    \label{fig:method}
\end{figure*}

\section{Theoretical framework}

To extract elastic constants accurately from first-principles calculations, particularly at finite temperatures, one must be precise about the definitions of strain and stress being utilized. Here, we outline the relevant thermodynamic measures and how they connect to outputs from density functional theory (DFT) and molecular dynamics (MD). 

\subsection{Strain and stress measures}
In continuum mechanics, the deformation of a crystal from its reference configuration is described by the deformation gradient tensor $\mathbf{D}$, with components $D_{ij} = \delta_{ij} + \partial u_i / \partial X_j$, where $\delta_{ij}$ is the Kronecker delta, $\mathbf{u}$ is the displacement vector, $\mathbf{X}$ is the reference position, and all the indices $i, j \in \{1, 2, 3\}$ correspond to the Cartesian spatial coordinates $x, y$, and $z$, respectively. For arbitrary deformations, the exact, frame-invariant measure of strain is the Green--Lagrange strain tensor:
\begin{equation}
\varepsilon_{ij} = \frac{1}{2} \Bigl(\sum_k D_{ki} D_{kj} - \delta_{ij}\Bigr).
\label{eq:lagrangian_strain}
\end{equation}

The choice of strain determines its work-conjugate stress measure. The true physical stress acting in the deformed configuration, and the quantity typically reported by DFT codes, is the Cauchy stress tensor $\sigma_{ij}$ (often denoted $\tau_{ij}$ in finite-strain theory). 

However, for constitutive modeling in the reference configuration, the natural stress measure is the second Piola--Kirchhoff stress tensor, $\mathbf{P}^{II}$ \cite{Klarbring2018}. This tensor is work-conjugate to the Green--Lagrange strain and measures forces and areas relative to the undeformed state:
\begin{equation}
P_{ij}^{II} = J \sum_{k,l} (D^{-1})_{ik} \sigma_{kl} (D^{-1})_{jl},
\label{eq:second_piola}
\end{equation}
where $J = \det(\mathbf{D})$ is the volume ratio.

\subsection{Helmholtz free energy and the small-strain limit}

The isothermal elastic constants are defined via the Taylor expansion of the Helmholtz free energy density $F/V$ with respect to the Green--Lagrange strain. For a stress-free reference state, the expansion to second order is:
\begin{equation}
\frac{F(\varepsilon)}{V} \approx \frac{F_0}{V} + \frac{1}{2}\sum_{i,j,k,l} c_{ijkl}\,\varepsilon_{ij}\varepsilon_{kl},
\label{eq:linear_elastic}
\end{equation}
where $F_0$ is the reference free energy, $V$ is the reference (undeformed) volume, and $c_{ijkl}$ are the second-order elastic constants. The corresponding stress--strain relationship is obtained by differentiation:
\begin{equation}
P_{ij}^{II} = \frac{1}{V}\frac{\partial F}{\partial \varepsilon_{ij}} = \sum_{k,l} c_{ijkl}\,\varepsilon_{kl}.
\label{eq:stress_strain_exact}
\end{equation}

Connecting this formal thermodynamic framework to first-principles calculations requires careful consideration, since DFT, as we have mentioned above, outputs the Cauchy stress $\sigma_{ij}$ rather than $P_{ij}^{II}$. However, in the linear-elastic regime (small strains, $|\partial u_i/\partial X_j| \ll 1$), the volume ratio $J \approx 1$ and $(D^{-1})_{ij} \approx \delta_{ij}$. Under these conditions, the Green--Lagrange strain reduces to the infinitesimal strain tensor, and all stress measures converge:
\begin{equation}
P_{ij}^{II} \approx \sigma_{ij}.
\label{eq:stress_convergence}
\end{equation}
This convergence justifies the direct use of the Cauchy stress obtained from DFT in the linear Hooke's law:
\begin{equation}
\sigma_{ij}=\sum_{k,l} c_{ijkl}\,\varepsilon_{kl}.
\label{eq:linear_stress_strain}
\end{equation}

\subsection{Finite-temperature formulation}

At finite temperature $T$ , the relevant thermodynamic potential is the Helmholtz free energy
\begin{equation}
F(T,V,\varepsilon_{ij})=-k_B T \ln Z(T,V,\varepsilon_{ij}),
\label{eq:helmholtz_definition}
\end{equation}
with canonical partition function
\begin{equation}
Z(T,V,\varepsilon_{ij})=\sum_\alpha \exp\!\left[-\beta E_\alpha(\varepsilon_{ij})\right],
\label{eq:partition_function}
\end{equation}
where $\beta=1/(k_B T)$, with $k_B$ representing the Boltzmann constant, and the sum runs over all accessible microstates $\alpha$. The thermodynamic stress is obtained from
\begin{equation}
\sigma_{ij}=\frac{1}{V}\left(\frac{\partial F}{\partial \varepsilon_{ij}}\right)_T.
\label{eq:thermodynamic_stress}
\end{equation}
Combining Eqs.~(\ref{eq:helmholtz_definition})--(\ref{eq:thermodynamic_stress}) yields
\begin{align}
\sigma_{ij}
&=-\frac{k_B T}{V}\frac{1}{Z}\frac{\partial Z}{\partial \varepsilon_{ij}} \nonumber\\
&=\frac{1}{V}\sum_\alpha \frac{\partial E_\alpha}{\partial \varepsilon_{ij}}\,
\frac{\exp(-\beta E_\alpha)}{Z} \nonumber\\
&=\frac{1}{V}\left\langle \frac{\partial E}{\partial \varepsilon_{ij}}\right\rangle.
\label{eq:ensemble_stress}
\end{align}

In an ergodic molecular-dynamics simulation, the time-averaged microscopic stress therefore coincides with the canonical ensemble average:
\begin{equation}
\sigma_{ij}^{\mathrm{MD}}=\frac{1}{t_s}\int_0^{t_s}\sigma_{ij}^{\mathrm{mic}}(t)\,dt,
\label{eq:md_stress}
\end{equation}
where $t_s$ is the time interval and
\begin{equation}
\sigma_{ij}^{\mathrm{mic}}(t)=\frac{1}{V}\frac{\partial E_\alpha(t)}{\partial \varepsilon_{ij}}.
\label{eq:microscopic_stress}
\end{equation}

This distinction is important. At zero temperature ($T = 0$~K), the free energy reduces to the ground-state total energy, and elastic constants can be obtained from either energy--strain or stress--strain relationships. At finite temperature, however, the time average of the internal energy from molecular dynamics is not, by itself, the Helmholtz free energy: the former represents $\langle E \rangle$, whereas $F=-k_B T \ln Z$ also contains the entropy contribution. A direct energy-based route therefore requires an appropriate free-energy evaluation or reconstruction, as in methods applied to post-processed \emph{ab initio} molecular-dynamics data, including TDEP-related approaches \cite{Hellman2011,Hellman2013,Knoop2024}.

The stress route provides a complementary and direct formulation. When the thermodynamic stress is obtained as an ensemble average, the finite-temperature elastic response can be extracted from its strain derivative without explicitly reconstructing the full free-energy surface. Thus, both free-energy--strain and stress--strain approaches are thermodynamically consistent when the corresponding thermodynamic quantities are evaluated correctly; the practical advantage of the present method is that it operates directly on the stress response. 

\subsection{Hooke's law in Voigt notation}

Owing to the minor and major symmetries of the elastic tensor, only 21 out of the 81 elements $c_{ijkl}$ are independent. In Voigt notation one makes the identifications $11 \rightarrow 1$, $22 \rightarrow 2$, $33 \rightarrow 3$, $23{=}32 \rightarrow 4$, $13{=}31 \rightarrow 5$, and $12{=}21 \rightarrow 6$. The stress tensor is then represented as a 6-component vector $\underline{\sigma}$, and the strain tensor as a 6-component vector $\underline{\epsilon}$, with the shear components multiplied by a factor of 2 (e.g., $\epsilon_4 = 2\varepsilon_{23}$).

Hooke's law then takes the compact matrix form:
\begin{equation}
\label{eq:hooks_law_matrix}
\underline{\sigma}=\underline{\underline{C}}\,\underline{\epsilon},
\end{equation}
where $\underline{\underline{C}}$ is the $6\times 6$ symmetric stiffness matrix containing the constants $c_{ij}$.

\section{Method}
\label{Method}

Several approaches have been developed to reduce the number of first-principles strain configurations required to determine the elastic tensor from stress--strain data. The most direct general construction applies each of the six independent strain components separately,
\begin{align*}
\underline{\epsilon} = {} & (\delta, 0, 0, 0, 0, 0)^T, \\
                 & (0, \delta, 0, 0, 0, 0)^T, \\
                 & \ldots, \\
                 & (0, 0, 0, 0, 0, \delta)^T,
\end{align*}
which is referred to as the \emph{All Single-Element Strain-Matrix Sets} (ASESS) in Ref.~\onlinecite{Liu2023}. This construction is general but redundant because it does not exploit symmetry and requires six separate strain configurations even when substantially fewer suffice.

A symmetry-reduced approach was introduced by Le Page and Saxe \cite{Page2002} and forms the basis of the conventional finite-difference stress--strain workflow implemented in the widely used Vienna Ab Initio simulation Package (VASP) \cite{Kresse1,Kresse2}. The corresponding strain sets are chosen from the symmetry of the crystal and the resulting stress--strain data are used to determine the independent elastic constants. This established scheme is efficient, but the published construction is presented through symmetry-specific strain sets rather than through a common generator that explicitly searches for a minimal uncoupled set.

The \emph{Universal Linear-Independent Coupling Strains} (ULICS) of Ref.~\cite{Yu2010} take a different route. The strain vectors are linearly independent in the six-dimensional strain space and contain no zero components, so several elastic constants contribute simultaneously to the resulting stress components. This can reduce the number of strained configurations; for example, two ULICS configurations suffice for a hexagonal crystal, compared with three in the Le Page--Saxe scheme \cite{Page2002}. The price is that the strained structure can lose additional symmetry, and the extraction of an individual constant relies on coupled stress equations rather than on a single stress component.

The \emph{Optimized High Efficiency Strain-Matrix Sets} (OHESS) introduced in Ref.~\cite{Liu2023} address the symmetry-preservation issue of ULICS by constructing more selective strain patterns. OHESS provides explicit, symmetry-specific sets that can reduce the number of configurations while avoiding unnecessary symmetry lowering. For example, four OHESS strains yield the thirteen independent elastic constants of a monoclinic crystal, compared with five ULICS strains \cite{Yu2010,Liu2023}. This reduction, however, is not achieved without limitations. In the tetragonal cases considered in Ref.~\cite{Liu2023}, the proposed strain sets do not directly determine all independent elastic constants: $c_{66}$ is not obtained from the stated strains and therefore requires an additional calculation. Thus, while OHESS can reduce the number of configurations compared with previous approaches, its prescribed strain sets do not always achieve the full reduction suggested by the nominal number of strains.

The present work takes a different perspective. Rather than prescribing a strain set for each symmetry, we formulate the construction directly in terms of the zero/nonzero pattern of the symmetry-reduced stiffness matrix. This makes it possible to generate admissible strains algorithmically and to separate the questions of \emph{construction} and \emph{minimality}. The resulting sets are uncoupled in the sense defined above, and their minimality is verified exactly by exhaustive enumeration of all admissible strain patterns. The number of applied configurations and the resulting extraction efficiency are summarized in Table~\ref{tab:calc_ratio}.

\begin{table*}[ht]
\centering
\caption{Ratios between the number of applied strain configurations (numerator) and the number of independent elastic constants obtained (denominator) for different crystal symmetries and computational methods. A lower ratio corresponds to higher computational efficiency. Cells highlighted by red frames indicate cases where previous methods are less efficient than the present work. 
}
\resizebox{\textwidth}{!}{%
\begin{tabular}{lccccccccc}
\hline
\textbf{Method} & \textbf{Triclinic} & \textbf{Monoclinic} & \textbf{Orthorhombic} & \textbf{Tetragonal-I} & \textbf{Tetragonal-II} & \textbf{Trigonal-I} & \textbf{Trigonal-II} & \textbf{Hexagonal} & \textbf{Cubic} \\
\hline
Ref.~\cite{Page2002}     & \fcolorbox{red}{white}{7/21} & \fcolorbox{red}{white}{5/13} & \fcolorbox{red}{white}{4/9} & \fcolorbox{red}{white}{3/7} & \fcolorbox{red}{white}{3/6} & \fcolorbox{red}{white}{3/7} & \fcolorbox{red}{white}{3/6} & \fcolorbox{red}{white}{3/5} & \fcolorbox{red}{white}{2/3} \\
Ref.~\cite{Yu2010} & 6/21 & \fcolorbox{red}{white}{5/13} & 3/9 & 2/7 & 2/6 & 2/7 & 2/6 & 2/5 & 1/3 \\
Ref.~\cite{Liu2023}      & 6/21 & 4/13 & 3/9 & \fcolorbox{red}{white}{2/6} & \fcolorbox{red}{white}{2/5} & 2/7 & 2/6 & 2/5 & 1/3 \\
Present work             & 6/21 & 4/13 & 3/9 & 2/7 & 2/6 & 2/7 & 2/6 & 2/5 & 1/3 \\
\hline
\end{tabular}
}
\label{tab:calc_ratio}
\end{table*}

\section{General algorithm for any crystal symmetry}
\label{Algorithm}

The purpose of the algorithm is to construct a minimum set of strain vectors whose associated stress equations are uncoupled for the independent elastic constants of a given crystal symmetry. In Voigt notation,
\begin{equation}
\sigma_i = \sum_{j=1}^{6} c_{ij}\, \epsilon_j ,
\end{equation}
where the symmetry of the crystal determines which elements of the stiffness matrix are identically zero and which nonzero elements represent independent or symmetry-equivalent elastic constants.

\subsection{Auxiliary matrix construction}

We define an \emph{auxiliary matrix} $\underline{\underline{A}}$ with entries
\begin{equation}
A_{ij} =
\begin{cases}
1, & c_{ij} \neq 0,\\
0, & c_{ij} \equiv 0,
\end{cases}
\end{equation}
so that the matrix records only the sparsity pattern of the stiffness tensor. The numerical magnitudes and signs of the elastic constants are deliberately discarded at this stage; only the possible couplings between stress and strain components are needed to construct the strain patterns.

\subsection{Reduction of symmetry-equivalent rows}

Rows corresponding to symmetry-equivalent diagonal elastic constants, for example $c_{ii}=c_{jj}$, need not be treated independently when constructing the strain patterns. Once one representative row from each equivalence class has been retained, the remaining rows from the same classes do not introduce additional independent information for the present construction. This follows from invariance of the elastic stiffness tensor under the symmetry operations of the crystal point group. A symmetry operation represented by a $3\times3$ orthogonal matrix $Q$ transforms the fourth-rank tensor according to $c'_{ijkl}=Q_{im}Q_{jn}Q_{kp}Q_{lq}c_{mnpq}$ (using Einstein summation convention), and tensor invariance imposes the relations that group symmetry-equivalent components into the same elastic constants, up to the sign conventions associated with shear components.

The equivalence need not correspond to a single symmetry operation that maps one coordinate axis directly onto another. For example, in a hexagonal crystal the equality $c_{11}=c_{22}$ follows from the combined constraints imposed by rotational symmetry about the $c$ axis. For the algorithm, the consequence is simply that one representative row is sufficient. We therefore construct a \emph{reduced auxiliary matrix} $\underline{\underline{A}}^{\,\mathrm{red}}$ by zeroing all but the smallest-index row within each set of rows with the equivalent diagonal elements. Rows whose diagonal element is not an independent constant, such as row 6 with $c_{66}=(c_{11}-c_{12})/2$ in hexagonal and trigonal crystals, are zeroed as well. The matrix remains $6\times6$ so that its row indices continue to correspond directly to stress components.

\subsection{Iterative strain construction algorithm}

The minimal set of uncoupled strain vectors is constructed iteratively as follows:

\begin{enumerate}
    \item Initialize an input auxiliary vector for iteration $k=1$, $\underline{B}^{(k)}_{\mathrm{in}}\in\{0,1\}^6$, by setting
    \[
    B^{(k)}_{\mathrm{in},i}=1
    \]
    for every strain component $i$ associated with at least one nonzero entry in row $i$ of $\underline{\underline{A}}^{(k)}=\underline{\underline{A}}^{\,\mathrm{red}}$, and zero otherwise.
    \item Compute
    \[
    \underline{v}=\underline{\underline{A}}^{(k)}\,\underline{B}^{(k)}_{\mathrm{in}}.
    \]
    The nonzero value $v_i$ counts the number of strain components that enter the retained equation for stress component $i$.
    \item If every nonzero component of $\underline{v}$ equals one, the current $\underline{B}^{(k)}_{\mathrm{in}}$ is accepted as $\underline{B}^{(k)}_{\mathrm{out}}$.
    \item Otherwise, identify the largest index $j_{\max}$ for which $v_{j_{\max}}>1$, temporarily (for this iteration only) set $B^{(k)}_{\mathrm{in},\ j_{max}} = 0$ and zero row $j_{\max}$ of $\underline{\underline{A}}^{(k)}$, recompute $\underline{v}$, and repeat the previous step until every nonzero component of $\underline{v}$ equals one. The resulting binary vector is $\underline{B}^{(k)}_{\mathrm{out}}$.
    \item Construct the $k$th strain vector as
    \begin{equation}
    \underline{\epsilon}^{(k)} =
    \begin{pmatrix}
    \delta_1^{(k)} B^{(k)}_{\mathrm{out},1} \\
    \delta_2^{(k)} B^{(k)}_{\mathrm{out},2} \\
    \vdots \\
    \delta_6^{(k)} B^{(k)}_{\mathrm{out},6}
    \end{pmatrix},
    \end{equation}
    where 
    the nonzero $\delta_j^{(k)}$ are independent strain amplitudes (numbered consecutively as $\delta_1, \delta_2, \ldots$ in the examples and tables).
    \item Undo the temporary changes of step 4. Then, form $\underline{B}^{(k+1)}_{\mathrm{in}}$ from $\underline{B}^{(k)}_{\mathrm{in}}$ by setting $\underline{B}^{(k+1)}_{\mathrm{in}, i}=0$ for all indices with $\underline{B}^{(k)}_{\mathrm{out}, i} =1$ and permanently zero the corresponding rows of $\underline{\underline{A}}^{(k)}$ to form $\underline{\underline{A}}^{(k+1)}$.
    \item Repeat, terminating at step $k$ when the reduced auxiliary matrix at step $k+1 $ becomes identically zero, indicating that no new independent information can be obtained. The resulting set of strains is the minimum strain set required for extraction. 

\end{enumerate}

A conceptual illustration of this iterative scheme is provided in Fig.~\ref{fig:method}.

\subsection{Interpretation of the method}

The construction can be formulated as a finite combinatorial covering problem over the sparsity pattern of the symmetry-reduced stiffness matrix. An admissible strain pattern produces stress equations in which no retained target equation contains more than one elastic constant. A collection of admissible strains is sufficient when their union covers every independent elastic constant.

The iterative procedure described above provides a deterministic greedy construction of such a collection. The purpose of the greedy step is not to replace the exact minimality test, but to generate a compact candidate set directly from the tensor sparsity. The formulation is closely related to finite set-cover constructions \cite{Wolsey2020}. Because the strain space is only six-dimensional, all admissible strains can subsequently be enumerated and their minimum number can be determined exactly. The stress relations not used directly for extraction remain available as redundant consistency checks.

\subsection{Exact verification of minimality}

The iterative construction is designed to identify a minimum strain set directly. Its minimality can be verified independently by exhaustive enumeration. For each elastic-symmetry type, all non-empty  strain patterns (subsets of the six Voigt components) can be examined, giving at most $2^6-1=63$ candidates. Retaining only those satisfying the uncoupling condition defined above reduces the problem to a finite covering problem that can be solved exactly. The minimum number of admissible strains is then the smallest number for which all independent elastic constants can be extracted from the corresponding strain set.

For every symmetry considered here, the set generated by the iterative construction attains this exact minimum. Thus, although the minimality of the iterative procedure is not established analytically, it is guaranteed for the present cases by exhaustive enumeration of all admissible strain sets.

The explicit operation of the algorithm is illustrated in the following section for Tetragonal-I symmetry. This example also shows how redundant stress relations emerge naturally from the same strained configurations and can be used as internal numerical checks.

\section{Application: Worked example for Tetragonal-I}

We select the Tetragonal-I symmetry (classes $4$, $\bar{4}$, and $4/m$, corresponding to space groups 75--88) to demonstrate the construction. This case is useful because the exact covering problem reveals a two-strain solution for all seven independent elastic constants. In the OHESS table of Ref.~\cite{Liu2023}, $c_{66}$ is not obtained from the listed two strains and therefore requires an additional calculation for the complete tensor; the same complete-tensor count is reflected in Table~\ref{tab:calc_ratio}.

First, we define the elastic tensor in Voigt notation:
\begin{equation}
\underline{\underline{C}}_{\mathrm{tet1}}=
\begin{pmatrix}
c_{11} & c_{12} & c_{13} & 0      & 0      & c_{16}\\
c_{12} & c_{11} & c_{13} & 0      & 0      & -c_{16}\\
c_{13} & c_{13} & c_{33} & 0      & 0      & 0\\
0      & 0      & 0      & c_{44} & 0      & 0\\
0      & 0      & 0      & 0      & c_{44} & 0\\
c_{16} & -c_{16}& 0      & 0      & 0      & c_{66}
\end{pmatrix}.
\label{eq:elastic_tetragonal1}
\end{equation}
This tensor contains seven independent elastic constants, four of which sit on the main diagonal. Mapping the non-zero elements to $1$ and the zero elements to $0$, we obtain the full auxiliary matrix $\underline{\underline{A}}_{\mathrm{tet1}}$:
\begin{equation}
\underline{\underline{A}}_{\mathrm{tet1}}=
\begin{pmatrix}
1 & 1 & 1 & 0 & 0 & 1\\
1 & 1 & 1 & 0 & 0 & 1\\
1 & 1 & 1 & 0 & 0 & 0\\
0 & 0 & 0 & 1 & 0 & 0\\
0 & 0 & 0 & 0 & 1 & 0\\
1 & 1 & 0 & 0 & 0 & 1
\end{pmatrix}.
\label{eq:aux_tetragonal1}
\end{equation} 

We then construct the reduced auxiliary matrix $\underline{\underline{A}}_{\mathrm{tet1}}^{\mathrm{red}}$ by zeroing out rows that are equivalent with respect to their diagonal elements, i.e., rows that contain the same elastic constants, allowing for differences in sign and multiplicity. Thus, in this case, row 2 provides no additional information beyond that contained in row 1, while row 5 provides no additional information beyond that contained in row 4.

\begin{equation}
\underline{\underline{A}}_{\mathrm{tet1}}^{\,\mathrm{red}}=
\begin{pmatrix}
1 & 1 & 1 & 0 & 0 & 1\\
0 & 0 & 0 & 0 & 0 & 0\\
1 & 1 & 1 & 0 & 0 & 0\\
0 & 0 & 0 & 1 & 0 & 0\\
0 & 0 & 0 & 0 & 0 & 0\\
1 & 1 & 0 & 0 & 0 & 1
\end{pmatrix}.
\label{eq:aux_tetragonal1_red}
\end{equation}

\textbf{Iteration 1:} We initialize the input auxiliary vector $\underline{B}^{1}_{\mathrm{in}}$ by marking a $1$ for every element $i$ if row $i$ has at least one non-zero entry in the reduced matrix:
\begin{equation}
\underline{B}^{1}_{\mathrm{in}} =
\begin{pmatrix}
1 \\ 0 \\ 1 \\ 1 \\ 0 \\ 1
\end{pmatrix}.
\end{equation}
Multiplying the reduced matrix by this vector gives us the number of coupled elastic constants for each stress component:
\begin{equation}
\underline{\underline{A}}_{\mathrm{tet1}}^{\,\mathrm{red}} \cdot \underline{B}^{1}_{\mathrm{in}} =
\begin{pmatrix}
3 \\ 0 \\ 2 \\ 1 \\ 0 \\ 2
\end{pmatrix}.
\end{equation}
Because several elements exceed $1$, the stress equations are coupled. Following the algorithm, we systematically set to zero the $\underline{B}^{1}_{\mathrm{in}}$ components corresponding to the highest-index component of $\underline{\underline{A}}_{\mathrm{tet1}}^{\,\mathrm{red}} \cdot \underline{B}^{1}_{\mathrm{in}}$ whose value exceeds $1$, and temporarily zero the corresponding rows of $\underline{\underline{A}}_{\mathrm{tet1}}^{\,\mathrm{red}}$; the products shown below are evaluated with these rows zeroed. We proceed in this way until no stress component depends on more than one strain parameter. In the present case, dropping component 6 and then component 3 yields a fully uncoupled vector:
\begin{equation}
\begin{pmatrix}
1 \\ 0 \\ 1 \\ 1 \\ 0 \\ 1
\end{pmatrix} \xrightarrow{\text{drop } 6}
\begin{pmatrix}
1 \\ 0 \\ 1 \\ 1 \\ 0 \\ 0
\end{pmatrix} \xrightarrow{\text{drop } 3}
\begin{pmatrix}
1 \\ 0 \\ 0 \\ 1 \\ 0 \\ 0
\end{pmatrix} = \underline{B}^{1}_{\mathrm{out}}.
\end{equation}
The corresponding products are
\begin{equation}
\begin{pmatrix}
3 & \\
0 & \\
2 & \\
1 & \\
0 & \\
\fcolorbox{red}{white}{$2$} & 
\end{pmatrix} \rightarrow
\begin{pmatrix}
2 & \\
0 & \\
\fcolorbox{red}{white}{$2$} & \\
1 & \\
0 & \\
0 & 
\end{pmatrix} \rightarrow
\begin{pmatrix}
1 & \\
0 & \\
0 & \\
1 & \\
0 & \\
0 & 
\end{pmatrix} = \underline{\underline{A}}_{\mathrm{tet1}}^{\,\mathrm{red}} \cdot \underline{B}^{1}_{\mathrm{out}}.
\label{eq:tet1_AB}
\end{equation}
Since in the final product in Eq.~\ref{eq:tet1_AB} all non-zero values are exactly $1$:
\begin{equation}
\underline{\underline{A}}_{\mathrm{tet1}}^{\,\mathrm{red}} \cdot \underline{B}^{1}_{\mathrm{out}} =
\begin{pmatrix}
1 \\ 0 \\ 0 \\ 1 \\ 0 \\ 0
\end{pmatrix},
\end{equation}
the first applied strain contains non-zero elements at indices 1 and 4:
\begin{equation}
\underline{\epsilon}^{\,I}_{\mathrm{tet1}} = \begin{pmatrix}
\delta_1^{I} \\ 0 \\ 0 \\ \delta_2^{I} \\ 0 \\ 0
\end{pmatrix},
\end{equation}
where $\delta_1^{I}$ and $\delta_2^{I}$ are independent small strain amplitudes chosen within the linear-response regime. In practical calculations, values of order $0.01$--$0.04$ are typically sufficient, with the appropriate range determined by convergence and linearity tests.

\textbf{Iteration 2:} We remove the previously used strain components (indices 1 and 4) to find the remaining required strains. The new input vector $\underline{B}^{2}_{\mathrm{in}}$ and the updated auxiliary matrix $\underline{\underline{A}}_{\mathrm{tet1}}^{(2)}$ (with rows 1 and 4 zeroed out) become:
\begin{equation}
\underline{B}^{2}_{\mathrm{in}} =
\begin{pmatrix}
0 \\ 0 \\ 1 \\ 0 \\ 0 \\ 1
\end{pmatrix}, \quad
\underline{\underline{A}}_{\mathrm{tet1}}^{(2)} =
\begin{pmatrix}
0 & 0 & 0 & 0 & 0 & 0\\
0 & 0 & 0 & 0 & 0 & 0\\
1 & 1 & 1 & 0 & 0 & 0\\
0 & 0 & 0 & 0 & 0 & 0\\
0 & 0 & 0 & 0 & 0 & 0\\
1 & 1 & 0 & 0 & 0 & 1
\end{pmatrix}.
\end{equation}
Multiplying these gives:
\begin{equation}
\underline{\underline{A}}_{\mathrm{tet1}}^{(2)} \cdot \underline{B}^{2}_{\mathrm{in}} =
\begin{pmatrix}
0 \\ 0 \\ 1 \\ 0 \\ 0 \\ 1
\end{pmatrix}.
\end{equation}
Since all non-zero elements are already equal to $1$, no further reduction is needed. The second strain vector contains elements at indices 3 and 6:
\begin{equation}
\underline{\epsilon}^{\,II}_{\mathrm{tet1}} = \begin{pmatrix}
0 \\ 0 \\ \delta_1^{II} \\ 0 \\ 0 \\ \delta_2^{II}
\end{pmatrix}.
\end{equation}

In summary, the minimal set consisting of two required strains for a Tetragonal-I crystal is:
\begin{equation}
\underline{\epsilon}_{\mathrm{tet1}}^{\,I}=
\begin{pmatrix}
\delta_1^{I}\\
0\\
0\\
\delta_2^{I}\\
0\\
0
\end{pmatrix},
\qquad
\underline{\epsilon}_{\mathrm{tet1}}^{\,II}=
\begin{pmatrix}
0\\
0\\
\delta_1^{II}\\
0\\
0\\
\delta_2^{II}
\end{pmatrix}.
\label{eq:delta_tetragonal1}
\end{equation}

Performing a first-principles calculation with strain $\underline{\epsilon}^{\,I}_{\mathrm{tet1}}$ yields five independent elastic constants directly from five distinct stress components:
\begin{equation}
\begin{matrix}
\sigma_{1}^{I}=c_{11}\delta_{1}^{I}, \\
\sigma_{2}^{I}=c_{12}\delta_{1}^{I}, \\
\sigma_{3}^{I}=c_{13}\delta_{1}^{I}, \\
\sigma_{4}^{I}=c_{44}\delta_{2}^{I}, \\
\sigma_{6}^{I}=c_{16}\delta_{1}^{I}.
\end{matrix}
\label{eq:c_tetragonal1_I}
\end{equation}
The second calculation with strain $\underline{\epsilon}^{\,II}_{\mathrm{tet1}}$ yields the remaining two independent constants:
\begin{equation}
\begin{matrix}
\sigma_{3}^{II}=c_{33}\delta_{1}^{II}, \\
\sigma_{6}^{II}=c_{66}\delta_{2}^{II}.
\end{matrix}
\label{eq:c_tetragonal1_II}
\end{equation}
All seven independent elastic constants are therefore obtained from single-stress-component relations using only two strained calculations. The same second strain also generates redundant relations:
\begin{equation}
\begin{matrix}
\sigma_{1}^{II}=c_{13}\delta_{1}^{II}+c_{16}\delta_{2}^{II}, \\
\sigma_{2}^{II}=c_{13}\delta_{1}^{II}-c_{16}\delta_{2}^{II},
\end{matrix}
\label{eq:c_tetragonal1_redundant}
\end{equation}
which provide a useful internal consistency check for the numerical accuracy of the DFT parameters.

\section{Strain matrices and extraction formulas for all symmetries}

For practical use, Tables~\ref{tab:summary_strains_1} and~\ref{tab:summary_strains_2} collect the strain matrices ($\mathbf{I}+\boldsymbol{\epsilon}$) and the corresponding extraction formulas for all elastic-symmetry types considered here. The matrices are the deformation gradients applied to the reference lattice vectors, and the $\delta_i$ are independent strain amplitudes. The indices in $\sigma$ and $c$ follow Voigt notation. Relations not used directly for extraction, but generated by the same deformations, provide redundant checks on the numerical stress response. 

\begin{table*}[ht]
\centering
\caption{Part I: Strain matrices and extraction formulas for Triclinic, Monoclinic, and Orthorhombic symmetries. The matrices shown are $\mathbf{I} + \boldsymbol{\epsilon}$.}
\label{tab:summary_strains_1}
\begin{tabular}{l l l}
\hline
\textbf{Symmetry} & \textbf{Strain Matrices} ($\mathbf{I} + \boldsymbol{\epsilon}$) & \textbf{Extraction Formulas} \\
\hline
\\
Triclinic (1--2) & $\epsilon^I$: \rule{0pt}{4ex} $\begin{pmatrix} 1+\delta_1 & 0 & 0 \\ 0 & 1 & 0 \\ 0 & 0 & 1 \end{pmatrix}$ & $\sigma_1^I=c_{11}\delta_1,\ \sigma_2^I=c_{12}\delta_1,\ \sigma_3^I=c_{13}\delta_1,\ \sigma_4^I=c_{14}\delta_1,\ \sigma_5^I=c_{15}\delta_1,\ \sigma_6^I=c_{16}\delta_1$ \\[6ex]
& $\epsilon^{II}$: \rule{0pt}{4ex} $\begin{pmatrix} 1 & 0 & 0 \\ 0 & 1+\delta_1 & 0 \\ 0 & 0 & 1 \end{pmatrix}$ & $\sigma_2^{II}=c_{22}\delta_1,\ \sigma_3^{II}=c_{23}\delta_1,\ \sigma_4^{II}=c_{24}\delta_1,\ \sigma_5^{II}=c_{25}\delta_1,\ \sigma_6^{II}=c_{26}\delta_1$ \\[6ex]
& $\epsilon^{III}$: \rule{0pt}{4ex} $\begin{pmatrix} 1 & 0 & 0 \\ 0 & 1 & 0 \\ 0 & 0 & 1+\delta_1 \end{pmatrix}$ & $\sigma_3^{III}=c_{33}\delta_1,\ \sigma_4^{III}=c_{34}\delta_1,\ \sigma_5^{III}=c_{35}\delta_1,\ \sigma_6^{III}=c_{36}\delta_1$ \\[6ex]
& $\epsilon^{IV}$: \rule{0pt}{4ex} $\begin{pmatrix} 1 & 0 & 0 \\ 0 & 1 & \frac{\delta_1}{2} \\ 0 & \frac{\delta_1}{2} & 1 \end{pmatrix}$ & $\sigma_4^{IV}=c_{44}\delta_1,\ \sigma_5^{IV}=c_{45}\delta_1,\ \sigma_6^{IV}=c_{46}\delta_1$ \\[6ex]
& $\epsilon^{V}$: \rule{0pt}{4ex} $\begin{pmatrix} 1 & 0 & \frac{\delta_1}{2} \\ 0 & 1 & 0 \\ \frac{\delta_1}{2} & 0 & 1 \end{pmatrix}$ & $\sigma_5^{V}=c_{55}\delta_1,\ \sigma_6^{V}=c_{56}\delta_1$ \\[6ex]
& $\epsilon^{VI}$: \rule{0pt}{4ex} $\begin{pmatrix} 1 & \frac{\delta_1}{2} & 0 \\ \frac{\delta_1}{2} & 1 & 0 \\ 0 & 0 & 1 \end{pmatrix}$ & $\sigma_6^{VI}=c_{66}\delta_1$ \\[6ex]
\hline
\\
Monoclinic (3--15) & $\epsilon^I$: \rule{0pt}{4ex} $\begin{pmatrix} 1+\delta_1 & 0 & 0 \\ 0 & 1 & \frac{\delta_2}{2} \\ 0 & \frac{\delta_2}{2} & 1 \end{pmatrix}$ & $\sigma_1^I=c_{11}\delta_1,\ \sigma_2^I=c_{12}\delta_1,\ \sigma_3^I=c_{13}\delta_1,\ \sigma_4^I=c_{44}\delta_2,\ \sigma_5^I=c_{15}\delta_1,\ \sigma_6^I=c_{46}\delta_2$ \\[6ex]
& $\epsilon^{II}$: \rule{0pt}{4ex} $\begin{pmatrix} 1 & \frac{\delta_2}{2} & 0 \\ \frac{\delta_2}{2} & 1+\delta_1 & 0 \\ 0 & 0 & 1 \end{pmatrix}$ & $\sigma_2^{II}=c_{22}\delta_1,\ \sigma_3^{II}=c_{23}\delta_1,\ \sigma_5^{II}=c_{25}\delta_1,\ \sigma_6^{II}=c_{66}\delta_2$ \\[6ex]
& $\epsilon^{III}$: \rule{0pt}{4ex} $\begin{pmatrix} 1 & 0 & 0 \\ 0 & 1 & 0 \\ 0 & 0 & 1+\delta_1 \end{pmatrix}$ & $\sigma_3^{III}=c_{33}\delta_1,\ \sigma_5^{III}=c_{35}\delta_1$ \\[6ex]
& $\epsilon^{IV}$: \rule{0pt}{4ex} $\begin{pmatrix} 1 & 0 & \frac{\delta_1}{2} \\ 0 & 1 & 0 \\ \frac{\delta_1}{2} & 0 & 1 \end{pmatrix}$ & $\sigma_5^{IV}=c_{55}\delta_1$ \\[6ex]
\hline
\\
Orthorhombic (16--74) & $\epsilon^I$: \rule{0pt}{4ex} $\begin{pmatrix} 1+\delta_1 & \frac{\delta_4}{2} & \frac{\delta_3}{2} \\ \frac{\delta_4}{2} & 1 & \frac{\delta_2}{2} \\ \frac{\delta_3}{2} & \frac{\delta_2}{2} & 1 \end{pmatrix}$ & $\sigma_1^I=c_{11}\delta_1,\ \sigma_2^I=c_{12}\delta_1,\ \sigma_3^I=c_{13}\delta_1,\ \sigma_4^I=c_{44}\delta_2,\ \sigma_5^I=c_{55}\delta_3,\ \sigma_6^I=c_{66}\delta_4$ \\[6ex]
& $\epsilon^{II}$: \rule{0pt}{4ex} $\begin{pmatrix} 1 & 0 & 0 \\ 0 & 1+\delta_1 & 0 \\ 0 & 0 & 1 \end{pmatrix}$ & $\sigma_2^{II}=c_{22}\delta_1,\ \sigma_3^{II}=c_{23}\delta_1$ \\[6ex]
& $\epsilon^{III}$: \rule{0pt}{4ex} $\begin{pmatrix} 1 & 0 & 0 \\ 0 & 1 & 0 \\ 0 & 0 & 1+\delta_1 \end{pmatrix}$ & $\sigma_3^{III}=c_{33}\delta_1$ \\[6ex]
\hline
\end{tabular}
\end{table*}

\begin{table*}[ht]
\centering
\caption{Part II: Strain matrices and extraction formulas for Tetragonal, Trigonal, Hexagonal, and Cubic symmetries.}
\label{tab:summary_strains_2}
\begin{tabular}{l l l}
\hline
\textbf{Symmetry} & \textbf{Strain Matrices} ($\mathbf{I} + \boldsymbol{\epsilon}$) & \textbf{Extraction Formulas} \\
\hline
\\
Tetragonal-I (75--88) & $\epsilon^I$: \rule{0pt}{4ex} $\begin{pmatrix} 1+\delta_1 & 0 & 0 \\ 0 & 1 & \frac{\delta_2}{2} \\ 0 & \frac{\delta_2}{2} & 1 \end{pmatrix}$ & $\sigma_1^I=c_{11}\delta_1,\ \sigma_2^I=c_{12}\delta_1,\ \sigma_3^I=c_{13}\delta_1,\ \sigma_4^I=c_{44}\delta_2,\ \sigma_6^I=c_{16}\delta_1$ \\[6ex]
& $\epsilon^{II}$: \rule{0pt}{4ex} $\begin{pmatrix} 1 & \frac{\delta_2}{2} & 0 \\ \frac{\delta_2}{2} & 1 & 0 \\ 0 & 0 & 1+\delta_1 \end{pmatrix}$ & $\sigma_3^{II}=c_{33}\delta_1,\ \sigma_6^{II}=c_{66}\delta_2$ \\[6ex]
\hline
\\
Tetragonal-II (89--142) & $\epsilon^I$: \rule{0pt}{4ex} $\begin{pmatrix} 1+\delta_1 & \frac{\delta_3}{2} & 0 \\ \frac{\delta_3}{2} & 1 & \frac{\delta_2}{2} \\ 0 & \frac{\delta_2}{2} & 1 \end{pmatrix}$ & $\sigma_1^I=c_{11}\delta_1,\ \sigma_2^I=c_{12}\delta_1,\ \sigma_3^I=c_{13}\delta_1,\ \sigma_4^I=c_{44}\delta_2,\ \sigma_6^I=c_{66}\delta_3$ \\[6ex]
& $\epsilon^{II}$: \rule{0pt}{4ex} $\begin{pmatrix} 1 & 0 & 0 \\ 0 & 1 & 0 \\ 0 & 0 & 1+\delta_1 \end{pmatrix}$ & $\sigma_3^{II}=c_{33}\delta_1$ \\[6ex]
\hline
\\
Trigonal-I (143--148) & $\epsilon^I$: \rule{0pt}{4ex} $\begin{pmatrix} 1+\delta_1 & 0 & 0 \\ 0 & 1 & 0 \\ 0 & 0 & 1 \end{pmatrix}$ & $\sigma_1^I=c_{11}\delta_1,\ \sigma_2^I=c_{12}\delta_1,\ \sigma_3^I=c_{13}\delta_1,\ \sigma_4^I=c_{14}\delta_1,\ \sigma_5^I=c_{15}\delta_1$ \\[6ex]
& $\epsilon^{II}$: \rule{0pt}{4ex} $\begin{pmatrix} 1 & 0 & 0 \\ 0 & 1 & \frac{\delta_2}{2} \\ 0 & \frac{\delta_2}{2} & 1+\delta_1 \end{pmatrix}$ & $\sigma_3^{II}=c_{33}\delta_1,\ \sigma_4^{II}=c_{44}\delta_2$ \\[6ex]
\hline
\\
Trigonal-II (149--167) & $\epsilon^I$: \rule{0pt}{4ex} $\begin{pmatrix} 1+\delta_1 & 0 & 0 \\ 0 & 1 & 0 \\ 0 & 0 & 1 \end{pmatrix}$ & $\sigma_1^I=c_{11}\delta_1,\ \sigma_2^I=c_{12}\delta_1,\ \sigma_3^I=c_{13}\delta_1,\ \sigma_4^I=c_{14}\delta_1$ \\[6ex]
& $\epsilon^{II}$: \rule{0pt}{4ex} $\begin{pmatrix} 1 & 0 & 0 \\ 0 & 1 & \frac{\delta_2}{2} \\ 0 & \frac{\delta_2}{2} & 1+\delta_1 \end{pmatrix}$ & $\sigma_3^{II}=c_{33}\delta_1,\ \sigma_4^{II}=c_{44}\delta_2$ \\[6ex]
\hline
\\
Hexagonal (168--194) & $\epsilon^I$: \rule{0pt}{4ex} $\begin{pmatrix} 1+\delta_1 & 0 & 0 \\ 0 & 1 & \frac{\delta_2}{2} \\ 0 & \frac{\delta_2}{2} & 1 \end{pmatrix}$ & $\sigma_1^I=c_{11}\delta_1,\ \sigma_2^I=c_{12}\delta_1,\ \sigma_3^I=c_{13}\delta_1,\ \sigma_4^I=c_{44}\delta_2$ \\[6ex]
& $\epsilon^{II}$: \rule{0pt}{4ex} $\begin{pmatrix} 1 & 0 & 0 \\ 0 & 1 & 0 \\ 0 & 0 & 1+\delta_1 \end{pmatrix}$ & $\sigma_3^{II}=c_{33}\delta_1$ \\[6ex]
\hline
\\
Cubic (195--230) & $\epsilon^I$: \rule{0pt}{4ex} $\begin{pmatrix} 1+\delta_1 & 0 & 0 \\ 0 & 1 & \frac{\delta_2}{2} \\ 0 & \frac{\delta_2}{2} & 1 \end{pmatrix}$ & $\sigma_1^I=c_{11}\delta_1,\ \sigma_2^I=c_{12}\delta_1,\ \sigma_4^I=c_{44}\delta_2$ \\[6ex]
\hline
\end{tabular}
\end{table*}
 
\section{Representative benchmark set}

To assess the practical consequences of the construction, we implemented the algorithm directly in VASP \cite{Kresse1,Kresse2} and compared it with the conventional finite-difference stress--strain workflow based on the Le Page--Saxe scheme \cite{Page2002}. We selected representative materials spanning all nine elastic-symmetry types. To isolate the effect of the strain-set construction, the compared calculations used identical projector augmented-wave (PAW) datasets and first-principles settings \cite{Blochl,Kresse2}. Specifically, the plane-wave cutoff was set to 1.4 times the largest \texttt{ENMAX} value in the corresponding PAW datasets, and the reciprocal-space sampling density \cite{Monkhorst1976} was 0.02~\AA$^{-1}$.
Exchange-correlation effects were treated using the Perdew--Burke--Ernzerhof (PBE) generalized-gradient approximation \cite{PBE}. In both workflows the strain amplitude was $\delta=0.015$ and each elastic constant was obtained from a two-point central difference between the $+\delta$ and $-\delta$ deformations (i.e. default \texttt{POTIM}$=0.015$ and \texttt{NFREE}$=2$),
 \begin{equation}
 c=\frac{\sigma_{+\delta}-\sigma_{-\delta}}{2\delta},
 \label{eq:central_diff}
 \end{equation}
with the stress component assigned to each constant as listed in Tables~\ref{tab:summary_strains_1} and~\ref{tab:summary_strains_2}, so that the two schemes differ only in the strain patterns applied. The results are summarized in  Table~\ref{tab:benchmark}.

The benchmark tests two distinct aspects of the method. First, the elastic constants obtained from the minimal uncoupled strain sets agree, within the numerical precision of the calculations, with those obtained from the conventional workflow. This shows that the reduction in the number of configurations does not alter the underlying linear stress--strain response. Because each target constant is extracted from a dedicated stress component, numerical errors in one stress component do not enter another constant through the algebraic solution of a coupled system.

Second, the wall-time reduction is not determined by strain count alone. The rightmost column of Table~\ref{tab:benchmark} reports the elapsed-time reduction under otherwise identical settings. The cost per strained configuration depends on the symmetry of the deformed structure and on the convergence behavior of the electronic calculation. Thus, a reduction in the number of configurations need not translate linearly into the same fractional reduction in wall time.

For the low-symmetry examples, including monoclinic and orthorhombic systems, the reduction in total wall time is comparatively modest ($7\%$ and $10\%$). The minimal construction combines several nonzero strain components within individual deformation matrices. These multi-component distortions can reduce the residual point-group symmetry of the strained structure more strongly than the corresponding symmetry-adapted strains in the conventional workflow \cite{Page2002}, increasing the size of the irreducible Brillouin zone (IBZ) and hence the number of irreducible $k$ points at fixed sampling density. This increases the cost per strained calculation and partly offsets the saving in the number of configurations.

For higher-symmetry systems, the wall-time reduction can exceed the reduction expected from configuration count alone. For example, decreasing the number of configurations from three to two for hexagonal systems, or from two to one for cubic systems, would correspond to nominal reductions of $33\%$ and $50\%$, respectively. The observed reductions are larger, reaching $55\%$ for Mg and $77\%$ for Al in the present benchmark. These results indicate that the computational cost of the individual strain calculations also changes substantially with the character of the deformation.

\begin{table*}[t]
\centering
\caption{Elastic constants (in GPa) obtained using the conventional finite-difference stress--strain implementation in VASP, based on the Le Page--Saxe scheme \cite{Page2002,Kresse1,Kresse2}, and using the present method. Representative materials spanning all crystal systems are shown; space groups (SG) are given in parentheses. All calculations use the PBE generalized-gradient approximation \cite{PBE} and exactly the same setups, such as the plane-wave energy cut-offs, the number of k-points etc. The \textbf{Exp.} column lists experimental single-crystal values used for comparison. The rightmost column reports the computational time saved, defined as $(t_{\mathrm{standard}}-t_{\mathrm{present}})/t_{\mathrm{standard}}\cdot100 \%$, where $t_{\mathrm{standard}}$ is the elapsed wall time for the conventional VASP workflow and $t_{\mathrm{present}}$ is the elapsed wall time for the present method under otherwise identical computational settings. Thus, positive values indicate the percentage of wall time saved relative to the standard implementation.}
\label{tab:benchmark}
\footnotesize
\renewcommand{\arraystretch}{1.12}
\setlength{\tabcolsep}{3.0pt}
\begin{tabular*}{\textwidth}{@{\extracolsep{\fill}} l l l l l l r @{}}
\hline\hline
\textbf{System} & \makecell[l]{\textbf{Material} \\ \textbf{(SG)}}
  & \textbf{Constants}
  & \textbf{Exp.}
  & \textbf{Standard} & \textbf{This work} & \makecell[r]{\textbf{Time} \\ \textbf{saved}} \\
\hline

Triclinic
  & NaAlSi$_{3}$O$_{8}$ (SG\,2)\cite{Brown2006}
  & \makecell[tl]{%
      $c_{11},\; c_{12},\; c_{13}$ \\[0pt]
      $c_{14},\; c_{15},\; c_{16}$ \\[0pt]
      $c_{22},\; c_{23},\; c_{24}$ \\[0pt]
      $c_{25},\; c_{26},\; c_{33}$ \\[0pt]
      $c_{34},\; c_{35},\; c_{36}$ \\[0pt]
      $c_{44},\; c_{45},\; c_{46}$ \\[0pt]
      $c_{55},\; c_{56},\; c_{66}$}
  & \makecell[tl]{%
    69.1, 34.0, 30.8 \\[0pt]
    5.1, $-$2.4, $-$0.9 \\[0pt]
    183.5, 5.5, $-$3.9 \\[0pt]
    $-$7.7, $-$5.8, 179.5 \\[0pt]
    $-$8.7, 7.1, $-$9.8 \\[0pt]
    24.9, $-$2.4, $-$7.2 \\[0pt]
    26.8, 0.5, 33.5}
  & \makecell[tl]{%
    63.1, 27.6, 28.3 \\[0pt]
    5.7, $-$5.1, $-$0.1 \\[0pt]
    162.1, 11.8, $-$3.6 \\[0pt]
    $-$7.8, $-$4.0, 163.2 \\[0pt]
    $-$5.5, 5.0, $-$8.9 \\[0pt]
    24.2, $-$1.4, $-$6.7 \\[0pt]
    26.9, 1.1, 33.6}
  & \makecell[tl]{%
    63.1, 27.6, 28.3 \\[0pt]
    5.7, $-$5.2, $-$0.1 \\[0pt]
    162.2, 11.8, $-$3.6 \\[0pt]
    $-$7.8, $-$4.0, 163.2 \\[0pt]
    $-$5.5, 5.0, $-$8.9 \\[0pt]
    24.2, $-$1.4, $-$6.7 \\[0pt]
    27.0, 1.1, 33.6}
  & $0\%$ \\[0pt]

Monoclinic
  & $\beta$-Ga$_{2}$O$_{3}$ (SG\,12)\cite{Adachi2018}
  & \makecell[tl]{%
      $c_{11},\; c_{12},\; c_{13}$ \\[0pt]
      $c_{15},\; c_{22},\; c_{23}$ \\[0pt]
      $c_{25},\; c_{33},\; c_{35}$ \\[0pt]
      $c_{44},\; c_{46},\; c_{55}$ \\[0pt]
      $c_{66}$}
  & \makecell[tl]{%
      242.8, 128.0, 160.0 \\[0pt]
      $-$1.6, 343.8, 70.9 \\[0pt]
      0.4, 347.4, 1.0 \\[0pt]
      47.8, 5.6, 88.6 \\[0pt]
      104.0}
  & \makecell[tl]{%
      204.1, 104.2, 109.5 \\[0pt]
      $-$16.5, 307.5, 62.8 \\[0pt]
      11.9, 307.3, 6.5 \\[0pt]
      44.9, 15.9, 64.6 \\[0pt]
      88.7}
  & \makecell[tl]{%
      205.4, 105.5, 109.4 \\[0pt]
      $-$16.7, 308.5, 62.5 \\[0pt]
      11.8, 305.6, 6.3 \\[0pt]
      45.5, 15.9, 64.6 \\[0pt]
      88.8}
  & $7\%$ \\[2pt]

Orthorhombic
  & TiSi$_{2}$ (SG\,70)\cite{Nakamura1994}
  & \makecell[tl]{%
      $c_{11},\; c_{12},\; c_{13}$ \\[0pt]
      $c_{22},\; c_{23},\; c_{33}$ \\[0pt]
      $c_{44},\; c_{55},\; c_{66}$}
  & \makecell[tl]{%
      317.5, 29.3, 38.4 \\[0pt]
      320.4, 86.0, 413.2 \\[0pt]
      112.5, 75.8, 117.5}
  & \makecell[tl]{%
      312.2, 27.2, 26.7 \\[0pt]
      313.3, 82.3, 399.6 \\[0pt]
      106.6, 73.1, 115.1}
  & \makecell[tl]{%
      312.2, 27.3, 26.7 \\[0pt]
      313.4, 82.5, 399.6 \\[0pt]
      106.3, 72.9, 115.3}
  & $10\%$ \\[2pt]

Tetragonal-I
  & CaMoO$_{4}$ (SG\,88)\cite{Wachtman1968CaMoO4}
  & \makecell[tl]{%
      $c_{11},\; c_{12},\; c_{13}$ \\[0pt]
      $c_{16},\; c_{33}$ \\[0pt]
      $c_{44},\; c_{66}$}
  & \makecell[tl]{%
      144.0, 64.8, 44.8 \\[0pt]
      $-$14.2, 126.0 \\[0pt]
      36.9, 46.1}
  & \makecell[tl]{%
      131.8, 52.6, 46.4 \\[0pt]
      $-$9.9, 114.3 \\[0pt]
      31.7, 38.7}
  & \makecell[tl]{%
      131.9, 52.7, 46.4 \\[0pt]
      $-$9.9, 114.2 \\[0pt]
      31.8, 38.7}
  & $29\%$ \\[2pt]

Tetragonal-II
  & TiO$_{2}$ (rutile, SG\,136)\cite{Wachtman1962Rutile}
  & \makecell[tl]{%
      $c_{11},\; c_{12},\; c_{13}$ \\[0pt]
      $c_{33},\; c_{44},\; c_{66}$}
  & \makecell[tl]{%
      266, 173, 136 \\[0pt]
      470, 124, 189}
  & \makecell[tl]{%
      256.6, 170.2, 147.6 \\[0pt]
      467.9, 113.9, 211.6}
  & \makecell[tl]{%
      256.8, 170.7, 147.8 \\[0pt]
      467.9, 113.9, 211.9}
  & $29\%$ \\[2pt]

Trigonal-I
  & CaMg(CO$_{3}$)$_{2}$ (SG\,148)\cite{Chen2006}
  & \makecell[tl]{%
      $c_{11},\; c_{12},\; c_{13}$ \\[0pt]
      $c_{14},\; c_{15}$ \\[0pt]
      $c_{33},\; c_{44}$}
  & \makecell[tl]{%
      204.1, 68.5, 45.8 \\[0pt]
      20.6, 6.7 \\[0pt]
      97.4, 39.1}
  & \makecell[tl]{%
      191.3, 65.9, 54.6 \\[0pt]
      16.7, 10.9 \\[0pt]
      106.5, 37.7}
  & \makecell[tl]{%
      192.6, 67.9, 55.4 \\[0pt]
      16.7, 10.9 \\[0pt]
      106.3, 38.1}
  & $27\%$ \\[2pt]

Trigonal-II
  & $\alpha$-SiO$_{2}$ (SG\,152)\cite{PabstGregorova2013,Heyliger2003}
  & \makecell[tl]{%
      $c_{11},\; c_{12},\; c_{13}$ \\[0pt]
      $c_{14},\; c_{33},\; c_{44}$}
  & \makecell[tl]{%
      86.6, 6.9, 13.6 \\[0pt]
      17.7, 106.7, 57.7}
  & \makecell[tl]{%
      79.2, $-$3.0, 6.9 \\[0pt]
      18.3, 87.0, 51.2}
  & \makecell[tl]{%
      79.7, $-$2.7, 7.4 \\[0pt]
      18.1, 87.7, 51.6}
  & $47\%$ \\[2pt]

Hexagonal
  & Mg (SG\,194)\cite{SlutskyGarland1957}
  & \makecell[tl]{%
      $c_{11},\; c_{12},\; c_{13}$ \\[0pt]
      $c_{33},\; c_{44}$}
  & \makecell[tl]{%
      59.7, 26.2, 21.7 \\[0pt]
      61.7, 16.4}
  & \makecell[tl]{%
      59.8, 26.7, 20.6 \\[0pt]
      62.5, 16.9}
  & \makecell[tl]{%
      60.2, 26.6, 20.5 \\[0pt]
      62.5, 17.0}
  & $55\%$ \\[2pt]

Cubic
  & Al (SG\,225)\cite{KammAlers1964}
  & $c_{11},\; c_{12},\; c_{44}$
  & 114.3, 61.9, 31.6
  & 114.2, 59.8, 34.2
  & 115.3, 59.1, 34.3
  & $77\%$ \\

\hline\hline
\end{tabular*}
\end{table*}

The data suggest that the additional wall-time savings observed for some high-symmetry systems are associated with the electronic-convergence behavior of the selected deformations. The conventional schemes can use strain patterns containing several simultaneously nonzero components, while the present construction introduces such combinations only when they are required to achieve the minimum number of configurations. In metallic systems, strongly multidimensional distortions may alter the electronic density and occupations more abruptly and thereby increase the number of self-consistent-field (SCF) iterations. The present strain sets often distribute the perturbation more selectively, which can reduce the cost per stress calculation. We regard this as a practical computational advantage of the generated strain patterns rather than as an intrinsic requirement of the algorithm.

The finite-temperature implications are direct. At finite temperature, obtaining the elastic response from stress--strain data requires statistically converged averages of thermodynamic stresses, so every additional strain configuration generally entails an additional molecular-dynamics trajectory or an additional ensemble of strained configurations. The benefit of minimizing the number of configurations therefore scales with the cost of the sampling itself. As a practical example, the present scheme was used in a first-principles study of body-centered cubic iron at inner-core conditions ($P=360$~GPa, $T=6600$~K) \cite{Belonoshko2022}. In that work, a 2000-atom supercell made stress averaging for a single strained configuration exceptionally expensive. For cubic symmetry, the present construction obtains all three independent constants from one strain set, whereas the conventional workflow requires two. Eliminating one complete strained simulation therefore removes a full stress-averaging workload in such an application.

\begin{figure*}[ht]
\centering
\includegraphics[width=\textwidth]{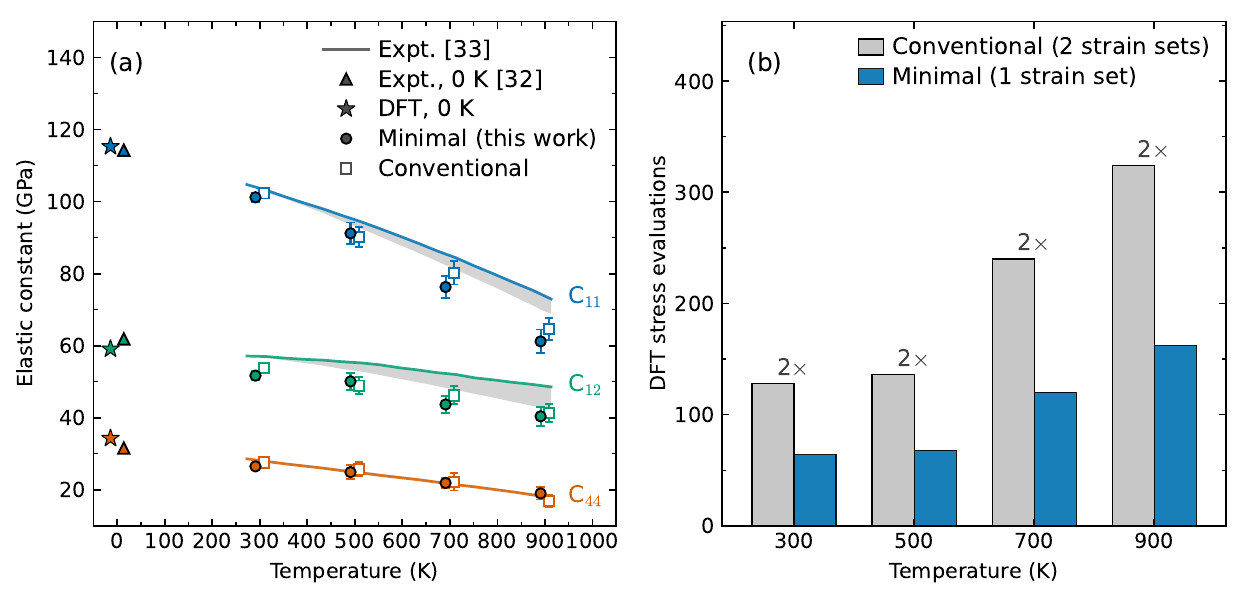}
\caption{Temperature dependence of the single-crystal elastic constants of fcc Aluminum. (a) $c_{11}$, $c_{12}$, and $c_{44}$ obtained from the present minimal scheme (filled circles) and from the conventional separate-strain scheme (open squares), each as the central difference of the density-functional thermodynamic stress averaged over molecular-dynamics configurations; error bars denote one standard error of the configurational mean and the two schemes are offset slightly in $T$ for clarity. Solid lines are the isothermal single-crystal data \cite{Tallon1979}, and stars mark the present $T=0$~K values. And, the grey region indicates the  spread of independent measurements \cite{Tallon1979, Gerlich1969}. The minimal and conventional results coincide within statistical error at every temperature. (b) Number of DFT stress evaluations required per temperature: the conventional scheme uses two strain sets where the minimal scheme uses one, and because the thermally disordered cells carry no global symmetry, the conventional approach is exactly twice as expensive ($2\times$) at every temperature.}
\label{fig:al_finiteT}
\end{figure*}

To verify that the minimal construction remains accurate when the stress is a thermodynamic average rather than a static response, we computed the temperature dependence of the single-crystal elastic constants of aluminum from $300$ to $900$~K. The canonical ensemble was sampled by \textit{ab initio} molecular dynamics in a $3\times3\times3$ conventional face-centered cubic (fcc) supercell ($108$ atoms) with a Langevin thermostat and $2$~fs timestep, accelerated with on-the-fly machine-learned force fields trained by Bayesian linear regression \cite{Jinnouchi2019}, as implemented in the \texttt{VASP} package.

At each temperature, the cell was first equilibrated for $20$~ps to its zero-pressure volume, from which the corresponding thermally expanded lattice constant $a(T)$ was obtained. The single minimal cubic strain pattern $\underline{\epsilon}^{I}_{\mathrm{cub}}=(\delta_1,0,0,\delta_2,0,0)^{T}$, with $\delta_1=\delta_2=\pm0.01$, was then applied to the thermally expanded lattice. 
The deformed cell was held rigid in both shape and volume while the ions were propagated for $60$~ps, with the first $10$~ps discarded as equilibration. The thermodynamic stress, Eq.~(\ref{eq:md_stress}), was evaluated with density functional theory on between $30$ and $88$ uncorrelated configurations per strain state, drawn at uniform intervals from the production trajectories, using a denser $8\times8\times8$ $\Gamma$-centered mesh in place of the $4\times4\times4$ mesh used during the dynamics. Each constant was then obtained from the same two-point central difference, Eq.~(\ref{eq:central_diff}), applied to the configuration-averaged stresses with $\sigma_2$ and $\sigma_3$ averaged before taking their difference since they are equal by symmetry but not configuration by configuration. Quoted uncertainties are one standard error of the configurational mean.

The results are shown in Fig.~\ref{fig:al_finiteT}(a). Across the entire temperature range, the minimal scheme follows the isothermal single-crystal measurements of Tallon and Wolfenden \cite{Tallon1979}, and the elastic constants obtained from the minimal and conventional strain sets agree within the statistical uncertainty. The residual underestimation of $c_{11}$ and $c_{12}$ at the highest temperatures is consistent with the thermal expansion predicted by PBE; the experimental single-crystal data also show noticeable scatter between independent measurements \cite{Tallon1979,Gerlich1969,KammAlers1964,Sutton1953}. Applying the conventional pair of single-component strains to the same  thermally expanded cells yields statistically indistinguishable elastic constants at every temperature. Thus, within the statistical precision of this benchmark, the uncoupled single-stress-component construction does not introduce an additional finite-temperature discrepancy relative to the conventional extraction.

The computational advantage is instead a reduction in the number of strained molecular-dynamics trajectories required per temperature. For cubic aluminum, the present scheme needs one strain set where the conventional workflow uses two, giving an exact factor-of-two reduction in the number of stress-averaging trajectories, as illustrated in Fig.~\ref{fig:al_finiteT}(b). Because each trajectory is accompanied by an ensemble stress evaluation, the saving scales with the full cost of the finite-temperature sampling rather than with the cost of a single static calculation. This makes strain-set minimality particularly valuable for large supercells and other calculations in which stress averaging dominates the computational expense.

\section{Summary}

We have presented a general algorithm for constructing a minimum set of strain vectors for second-order elastic-constant extraction from first-principles stress--strain calculations. The construction uses only the symmetry-reduced zero/nonzero structure of the elastic stiffness tensor, represented by an auxiliary binary matrix, and converts the choice of strain configurations into a finite combinatorial covering problem. The generated strain sets are \emph{uncoupled} in the sense that each elastic constant targeted for extraction is obtained from a single stress component, while additional relations from the same deformation can be retained as internal consistency checks.

A key feature of the six-component strain space is that the complete set of admissible strain patterns can be enumerated explicitly. We use this finite search to verify the exact minimum number of uncoupled configurations for all elastic-symmetry types spanning the 230 crystallographic space groups. The resulting sets reproduce established high-efficiency constructions where they are already minimal and provide a smaller complete set where an additional configuration is otherwise required.

The benchmark calculations show that the minimal construction reproduces the conventional finite-difference elastic constants within numerical precision while reducing the number of strained calculations. The wall-time reduction depends not only on the number of configurations but also on the electronic and Brillouin-zone cost of each deformation, and the present strain sets can therefore yield either moderate or substantially larger savings than the configuration count alone would suggest. At finite temperature, where every additional strain can require an additional stress-averaging trajectory, the benefit of minimality becomes even more direct. The algorithm is thus suitable for automated first-principles workflows, large-supercell calculations, and finite-temperature applications in which the cost of sampling the thermodynamic stress is significant.

\section{Data and Code Availability}

The data supporting the findings of this work are available from the corresponding author upon reasonable request. The patch for VASP used in this work is also available from the corresponding author, upon reasonable request subject to the license restrictions of the underlying software.

\vspace{-1.5em}
\begin{acknowledgments}
S.I.S. acknowledges support from the Swedish Research Council (VR) (2023-05247). 
U.S. acknowledges funding support from the Swedish e-Science Research Centre (SeRC). 
Computational resources were provided by the National Academic Infrastructure for Supercomputing in Sweden (NAISS), funded by the Swedish Research Council.
\end{acknowledgments}

\bibliographystyle{apsrev4-2}
\bibliography{references.bib}
\end{document}